%% file: main.tex
\documentclass[sigconf]{acmart}
\renewcommand\footnotetextcopyrightpermission[1]{} 
\usepackage[linesnumbered,ruled]{algorithm2e}
\title{Using Retriever Augmented Large Language Models for Attack Graph Generation}
\author{Renascence Tarafder Prapty}
\affiliation{University of California Irvine}
\email{rprapty@uci.edu} 
\author{Ashish Kundu} 
\affiliation{Cisco Research}
\email{ashkundu@cisco.com} 
\author{Arun Iyengar}
\affiliation{Cisco Research}
\email{ariyenga@cisco.com} 

\newcommand{\system}{{\ensuremath{\sf{CrystalBall}}\xspace}}
\begin{document}
\input{content/00_abstract}
\maketitle
\input{content/01_introduction}
\input{content/02_background}
\input{content/03_proposed_system}
\input{content/04_implementation}

\input{content/05_result}

\input{content/06_related_work}
\input{content/07_conclusion}
\bibliographystyle{unsrt}
\bibliography{reference}
\input{content/08_appendix}
\end{document}

%% file: content/00_abstract.tex
\begin{abstract}
As the complexity of modern systems increases, so does the importance of assessing their security posture through effective vulnerability management and threat modeling techniques. One powerful tool in the arsenal of cybersecurity professionals is the attack graph, a representation of all potential attack paths within a system that an adversary might exploit to achieve a certain objective. Traditional methods of generating attack graphs involve expert knowledge, manual curation, and computational algorithms that might not cover the entire threat landscape due to the ever-evolving nature of vulnerabilities and exploits. This paper explores the approach of leveraging large language models (LLMs), such as ChatGPT, to automate the generation of attack graphs by intelligently chaining Common Vulnerabilities and Exposures (CVEs) based on their preconditions and effects. It also shows how to utilize LLMs to create attack graphs from threat reports.
\end{abstract}

%% file: content/01_introduction.tex
\section{Introduction}
Attack graphs provide a comprehensive view of multiple attack vectors that a malicious actor could exploit to compromise system security. They are crucial tools for security analysts, enabling the understanding and visualization of potential attack paths in a system. Traditionally, constructing attack graphs has been a largely manual and time-consuming process requiring extensive expertise in cybersecurity. Automated techniques exist but they often rely on static rules or heuristics which cannot adapt to the ever-changing landscape of vulnerabilities and threats. Some existing approaches \cite{al2019a2g2v,sheyner2002automated,ou2006scalable} rely on static formal definitions and model checking algorithms to generate attack graphs. However, these methods are domain-specific and cannot be applied to emerging attack vectors. Furthermore, they require manual input of vulnerability information, which is not desirable since new vulnerabilities are continually discovered, and attack graphs should be updated promptly with this information. A new approach \cite{payne2019secure,bezawada2019agbuilder,husari2017ttpdrill} for automatic attack graph generation involves using natural language processing (NLP) techniques and machine learning models. This method uses NLP to extract different properties from vulnerability information found in a public database called the Common Vulnerabilities and Exposures (CVE). The extracted properties are utilized to create attack paths by chaining different CVEs. However, traditional machine learning models require an extensive training phase, which is a non-trivial task.

In recent years, advances in machine learning have presented new avenues for solving complex problems. Large language models (LLMs) like GPT-4, trained on diverse data sources, have demonstrated significant capabilities in natural language understanding and generation. These models have been applied successfully across a variety of domains, including translation, summarization, and question-answering among others. A natural question is how LLMs can be applied to the cybersecurity domain, specifically for generating attack graphs.

The aim of this paper is to investigate the potential of using large language models such as ChatGPT for automating the generation of attack graphs. Our approach leverages LLM capabilities to understand and chain Common Vulnerabilities and Exposures (CVEs) based on their preconditions and postconditions. By interpreting CVE descriptions and associated metadata, LLMs can generate links between vulnerabilities, offering a dynamic way to visualize possible attack vectors. In addition, this paper explores using LLMs for generating attack graphs based on textual threat reports, which are often rich sources of data but require manual analysis to transform into actionable insights.

Our work makes several contributions:
\begin{itemize}
    \item We present \system{}, a novel method for automated attack graph generation using retriever-augmented large language models. \system{} is designed to automatically generate attack graphs in a scalable manner. \system{} is able to handle significant amounts of unstructured text data as well as structured data.
    \item To that purpose, we design a new retriever model using retriever augmented generation technique (RAG) for correct and precise extraction of relevant CVEs based on the system information. 
    \item We develop a semantic search capable structured database to support the retriever.
    \item  We compare the performances of different LLMS for generating attack graphs.
    \item We evaluate the performances of different LLMS on creating attack graphs from threat reports as well.
\end{itemize}

This research not only offers a promising avenue for improving the efficiency and accuracy of attack graph generation but also provides insights into the broader applicability of large language models in cybersecurity.

%% file: content/02_background.tex
\section{Background}
\subsection{CVE}
Common Vulnerabilities and Exposures (CVE) \cite{cve} is a database of publicly disclosed information security issues which is maintained by the MITRE corporation \cite{mitre}. Each vulnerability entry of the database is also referred to as a CVE. All published CVEs are available in JSON Format in a github repository \cite{cve_repo}. Important properties of a CVE are CVE ID, State, and Description. CVE ID is a unique, alphanumeric identifier that uniquely identifies a specific vulnerability (CVE) from the database. CVE record or description is the descriptive data about a Vulnerability associated with a CVE ID. 

\begin{figure}[H]
    \centering
    \includegraphics[width=\linewidth]{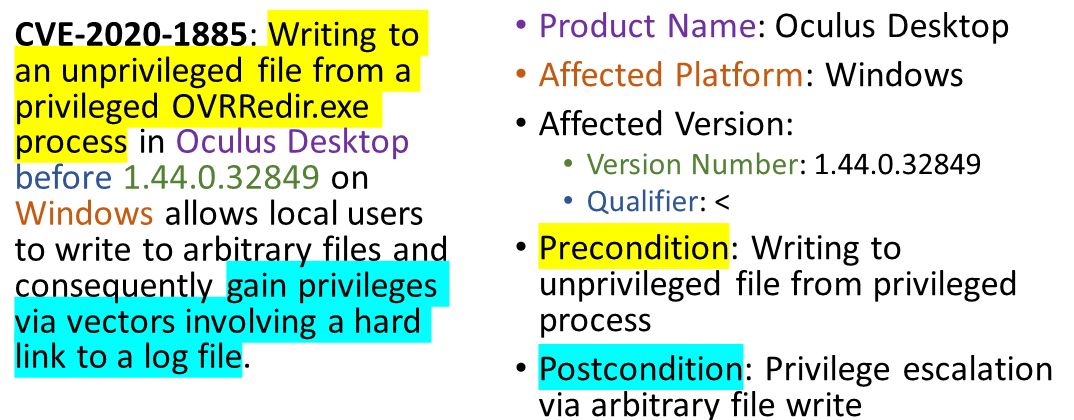}
    \caption{Semi-structured format of a CVE record description}
    \label{fig: cve description}
\end{figure}

CVE record has a semi-structured format. It represents a natural language description of the vulnerability. However, the description is bound to contain certain information i.e. product name, affected platform, affected version name etc. It also contains precondition and postcondition of the vulnerability. Figure \ref{fig: cve description} illustrates its semi-structured nature.

\subsection{Attack Graph}
\begin{figure*}
    \centering
    \includegraphics[width=\textwidth]{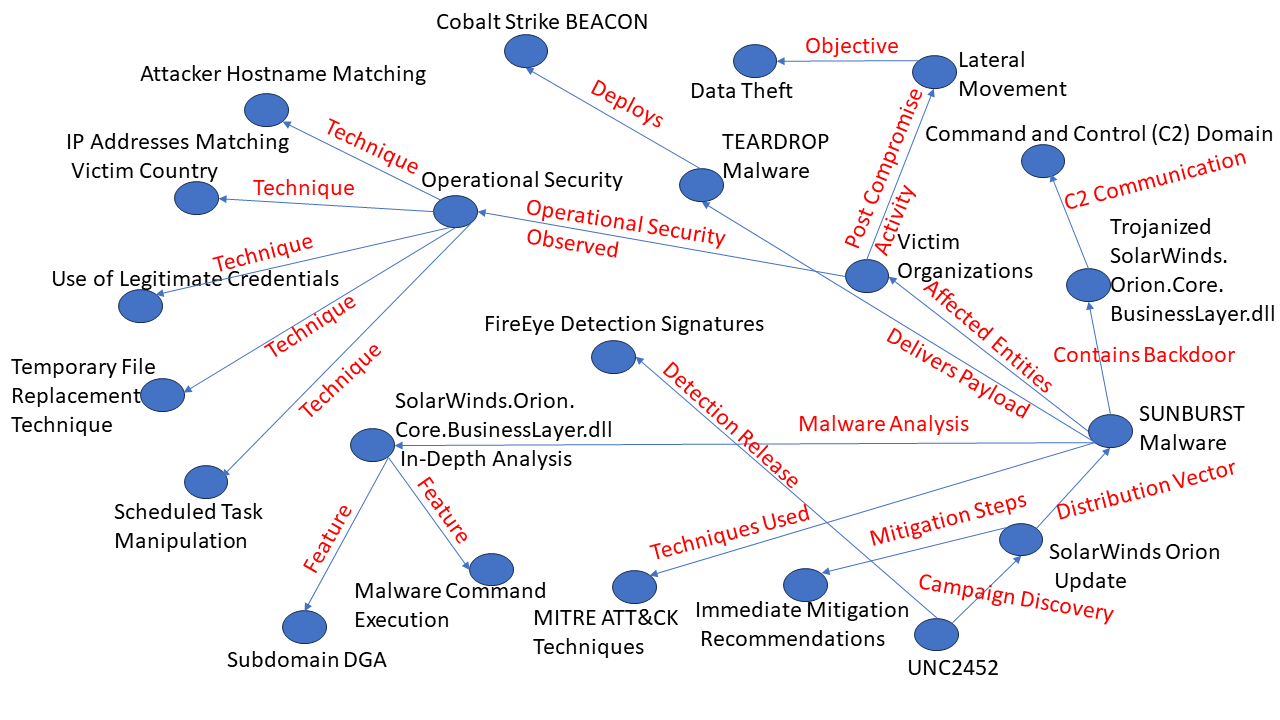}
    \caption{Atttack Graph constructed from threat report of SolarWinds Supply Chain Compromise Incident}
    \label{fig: solarwinds attack}
\end{figure*}

An attack graph is a model for the representation of all potential attack paths in a system. It supports a visualization model and shows how attackers can exploit different vulnerabilities, and chain the vulnerabilities together to achieve attack goals. Attack graphs are helpful in understanding the potential attack paths attackers can take and prioritizing security efforts. Figure \ref{fig: solarwinds attack} shows an example attack graph. The prompt for this graph is available at Appendix \ref{appendix: SolarWinds full}.

%% file: content/03_proposed_system.tex
\section{Proposed System}
\begin{figure}[H]
    \centering
    \includegraphics[width=\linewidth]{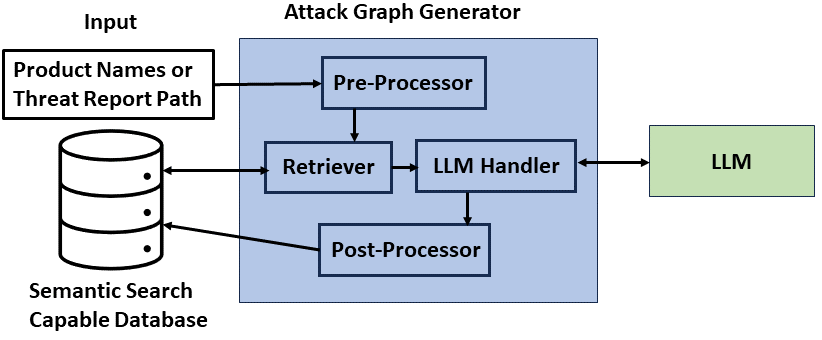}
    \caption{Proposed System Model}
    \label{fig: system model}
\end{figure}

We propose retriever-augmented generation using large language models(LLM) to generate attack graphs. Figure \ref{fig: system model} shows an overview of the proposed system. There are four main components in the system:

\begin{itemize}
    \item \textbf{Input}: Users can utilize our system to create attack graphs from either threat reports or CVE descriptions. In the case of CVE descriptions, users provide the target product/package name in the query. Retriever-augmented generation is used to obtain CVEs related to a user's query. Then these CVEs are used to create context for the prompt to the LLM. In case of a threat report, a path to the threat report text file is given in the query.
    \item \textbf{Database}: The proposed system utilizes a relational database that stores CVE records, related metadata, and generated attack graphs. This database can be searched using semantic similarity. Details about this semantic search process is available in Section \ref{sec: retriever}
    \item  \textbf{Attack Graph Generator}: This is the main component of the system which is responsible for handling user requests. It communicates with database to create prompts and with LLMs to obtain actual attack graphs. This module is described in detail later in the section.
    \item \textbf{LLM}: LLM is used as a black box in the system. It is responsible for creating the attack graphs from the information provided in the prompt.
\end{itemize}
 
Next we describe Attack Graph Generator module in details. It has four main parts: 1. Pre-Processor, 2. Retriever, 3. LLM Handler, 4. Post-Processor. 

\subsection{Pre-Processor}
Pre-Processor is responsible for processing the user input. In case of threat reports, Pre-Processor reads the report text using the input path and passes it as context to the LLM handler. On the other hand, when product names are given as queries to build an attack graph from CVE descriptions, Pre-Processor obtains the embeddings of each product name and provides these embeddings to the Retriever.
\subsection{Retriever} \label{sec: retriever}
Retrievers use either keyword matching or cosine similarity between embeddings from CVE information and user queries to find relevant CVEs. Special attention is required in determining which information from a CVE to use for embeddings. Traditional methods use embeddings of a whole CVE description. However, this results in low cosine similarity between the cve and query embedding. We observer that cosine similarity is inversely proportional to the length of CVE descriptions. If we keep the threshold of cosine similarity high for a CVE to be considered relevant, we risk missing related CVEs with long descriptions. On the other hand, choosing a low cosine similarity threshold results in extracting CVEs which are not actually related. 

Another issue is the granularity of the selection criteria. Some vulnerabilities might affect products of one specific platform only, e.g. a product running on Linux but not on Windows. A similar situation exists for the version number. Only specific versions of the product may be vulnerable. Moreover, a user may be interested in specific types of vulnerabilities only such as buffer overflow or remote code execution. Neither keyword search nor embedding from the whole description facilitates these fine tuned retrievals. 

\noindent \textbf{Developing the Retriever}
We propose a retriever model which is able to facilitate the above mentioned fine tune selection parameters. One key challenge is obtaining product name, problem type, platform and version properties for each CVE. CVE information available from the Mitre website is semi structured in json format. It has properties such as affected product/package name, problem type and version. However, values for these properties are often missing. No property is available for platform. Nevertheless all of these properties are usually included in the natural language text of CVE descriptions. Our proposed model utilizes these characteristics in an interesting and innovative way. 

\begin{figure}
    \centering
    \includegraphics[width=0.5\textwidth]{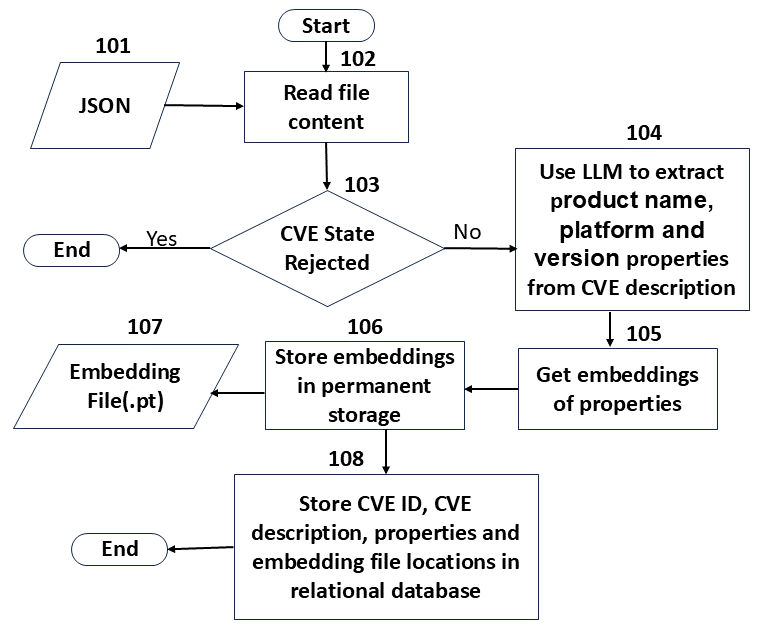}
    \caption{Retriever Development Phase}
    \label{fig:preprocessing}
\end{figure}

During the preprocessing phase, the LLM is used to extract the relevant properties from the description. Then the embedding for each property is created and stored in permanent storage. Finally these properties along with the storage location for corresponding embeddings are saved in multiple relational database files. Algorithm \ref{alg: preprocessing } provides the pseudocode for this process. Figure \ref{fig:preprocessing} shows the steps taken during development of retriever model.

\begin{algorithm}
    \SetKwInOut{Input}{Input}
    \SetKwInOut{Output}{Output}

    \Input{Path to CVE Json File P}
    Initialize LLM Client LC\\
    Initialize Embedding Model EM\\
    Initialize Database Client D\\
    cve\_json = Read File (P)\\
    state = cve\_json[``cveMetadata''][``state'']\\
    \eIf{state == `rejected'}{return}{
    cve\_id = cve\_json[``cveMetadata''][``cveId'']\\
    info = cve\_json[``containers''][``cna'']\\
    description = info[``descriptions''][0][``value'']\\
    \tcp{Table name followed by values to be inserted in that table}
    D.store(CVE\_INFO, cve\_id, description)\\
    
    prompt = ``provide the affected product name, platform and version if present using a key called ProductInfo, platform using a key called Platform, the problem type using a key called ProblemType in a single json from vulnerability information given below. ProductInfo should have properties ProductName and Version. Version should have properties VersionNumber and Qualifier whose 
    value would be <=, >=, == etc. '' + description \\
    result = LC.call(prompt)\\
    
    product\_name = result[ProductName]\\
    product\_name\_vector = EM.process(product\_name)\\
    product\_name\_vector\_file = cve\_id + ``\_product.pt''\\
    \tcp{Saves the vector in the storage location specified by vector file}
    save(product\_name\_vector, product\_name\_vector\_file)\\
    product\_id = D.store(PRODUCT\_INFO, cve\_id, product\_name, product\_name\_vector\_file)\\
    version = result[``ProductName''][``Version'']\\
    version\_number = version[``VersionNumber'']\\
    qualifier = version[``Qualifier'']\\
    D.store(VERSION\_INFO, product\_id, version\_number, qualifier)\\
    
    problem\_type = result[ProblemType]\\
    problem\_type\_vector = EM.process(problem\_type)\\
    problem\_type\_vector\_file = cve\_id + ``\_problem\_type.pt''\\
    save(problem\_type\_vector, problem\_type\_vector\_file)\\
    D.store(PROBLEM\_TYPE, cve\_id, problem\_type, problem\_type\_vector\_file)\\
    
    platform = result[Platform]\\
    platform\_vector = EM.process(platform)\\
    platform\_vector\_file = cve\_id + ``\_platform.pt''\\
    save(platform\_vector, platform\_vector\_file)\\
    D.store(PLATFORM, cve\_id, platform, platform\_vector\_file)\\
}
    \caption{Preprocess CVE (P)
    \label{alg: preprocessing }
    }
\end{algorithm}

\noindent \textbf{Getting Context Using Retriever}
While calling LLM to create attack graphs from vulnerability information of a system, we use Retriever to build relevant context. Algorithm \ref{alg: get context } provides the pseudocode of this process.

User provides the Product/Package names which make up the target system in a query to the Retriever. Retriever obtains embeddings for each name in the query from Pre-Processor. All product names and corresponding embedding locations are fetched from the database. Embedding of each product name is loaded in memory from permanent storage, and cosine similarity between this product name embedding and query name embedding is calculated. If the calculated similarity is greater than the defined threshold, corresponding CVE is considered relevant and its description in included in the context. While adding new CVE description to the context, the number of token in the context is calculated to check if the context is within the number of tokens allowed by LLM. The same steps are repeated for Platform information too. Our prototype Retriever model uses only product name and platform properties for retrieval. However, it can be easily extended for problem type and version properties as well. 

Our proposed Retriever is uniquely qualified to take advantage of semi structured nature of CVE information. It uses relational database to handle structured part i.e. fixed number of properties while using cosine similarity to handle natural language part i.e. using semantic similarity to match values for a particular property.

\begin{algorithm}
    \SetKwInOut{Input}{Input}
    \SetKwInOut{Output}{Output}

    \Input{User Query Q}    \Output{Context C which includes CVEs related to Q}
    Initialize Embedding Model EM\\
    Initialize Database Client D\\
    query\_vector = EM.process(Q)\\
    \tcc{minimum cosine similarity value allowed between query\_vector and a product\_vector or platform\_vector}
    min\_similarity = 0.68 \\
    \tcp{maximum number of tokens allowed in the context, can be configured}
    context\_tokens\_per\_query = 3750\\
    C = ``"\\
    relevant\_cves = set()\\
    retrieved\_products = D.query(PRODUCT\_INFO) \\
    \For{product in retrieved\_products}{
        product\_name\_vector\_file = product[3]\\
        cve\_id = product[1]\\
        product\_name\_vector = load(product\_name\_vector\_file)\\
        score = calculate\_cosine\_similarity(query\_vector, product\_name\_vector)\\
        \eIf{score > min\_similarity}{
        relevant\_cves.add(cve\_id)\\
        }
        {
        \tcp{do not include embedding in context due to low cosine similarity}
            continue 
        }
    }
    retrieved\_platforms = D.query(PLATFORM) \\
    \For{platform in retrieved\_platforms}{
        platform\_vector\_file = platform[3]\\
        cve\_id = platform[1]\\
        platform\_vector = load(platform\_vector\_file)\\
        score = calculate\_cosine\_similarity(query\_vector, platform\_vector)\\
        \eIf{score > min\_similarity}{
        relevant\_cves.add(cve\_id)\\
        }
        {
        \tcp{do not include embedding in context due to low cosine similarity}
            continue 
        }
    }
    retrieved\_cves = D\_query(CVE\_INFO, relevant\_cves)\\
    \For{cve in retrieved\_cves}{
    description = cve[1]\\
        num\_tokens = Calculate Num Token (C + ``---" + description)\\
        \eIf{num\_tokens < context\_tokens\_per\_query }{
        C = C+``---" + description
        }{
            break \tcp{C reached maximum token limit allowed in prompt}
        }
    }
  return C
\caption{Get Context (Q)}
\label{alg: get context }
\end{algorithm}

\subsection{LLM Handler}
\begin{figure*}
    \centering
    \includegraphics[width=\textwidth]{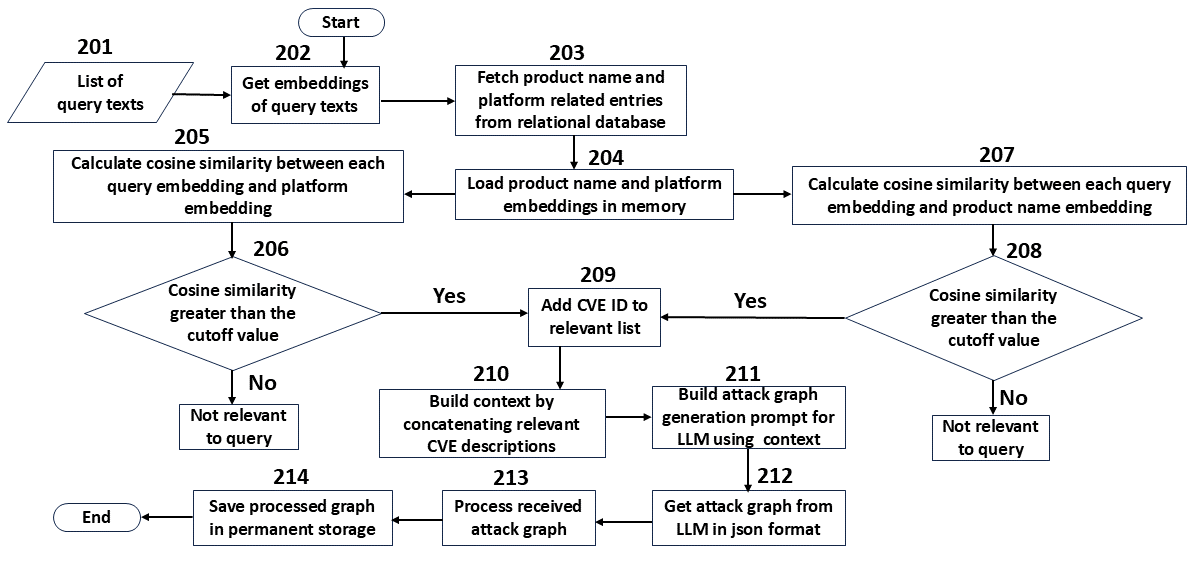}
    \caption{Attack Graph Generation Steps}
    \label{fig:generation}
\end{figure*}

After obtaining relevant context from Retriever, LLM Handler builds the prompt to LLM by adding a common prefix to the context. This common prefix includes some instructions for LLM such as the format of the output and how context should be treated. Then LLM Handler calls the LLM and receives the answer from it. Algorithm \ref{alg: generate attack graph } provides the pseudocode for this process. Figure \ref{fig:generation} presents the whole picture of attack graph generation for a system.

\begin{algorithm}
    \SetKwInOut{Input}{Input}
    \SetKwInOut{Output}{Output}

    \Input{User Query Q which includes product names}    \Output{Attack Graph G}
        context = Get Context(Q)\\
        prompt\_common = ``Create an attack graph in json format using nodes and edges from the vulnerability information given below. Chain the vulnerabilities if applicable. Vulnerabilities can be chained if precondition of one vulnerability is similar to the post condition of another vulnerability. Vulnerability with matching postcondition should be ahead in the chain. ''\\
        prompt = prompt\_common + context\\
        Initialize LLM Client LC\\
        answer = LC.call(prompt)\\
        G = process\_graph(answer)\\
        graph\_file\_name = ``graph-'' + timestamp + ``.json''\\
        save(G, graph\_file\_name)\\
        show(G)\\
      return G
    \caption{Generate Attack Graph (Q)
    \label{alg: generate attack graph }
    }
\end{algorithm}

\subsection{Post-Processor}
The answer from LLM is processed by Post-Processor. It extracts the attack graph from the answer, saves it in database and shows the graph to the user.

%% file: content/04_implementation.tex
\section{Implementation}
\system is implemented using Python.

\noindent \textbf{Pre-Processor}: Standard file I/O operations are used to read threat reports and CVE JSON files. Additionally, json package is used to process CVE JSON files. Pretrained facebook/contriever-msmarco model from Hugging Face\cite{huggingface} is used to obtain embeddings.

\noindent \textbf{Database:} We use SQLite as the relational database and utilize sqlite3 package to communicate with it.

\noindent \textbf{Retriever}: Retriever uses \textbf{torch} package to store embeddings in ``.pt'' vector files during the development phase and load embeddings into memory from ``.pt'' files during attack graph generation phase. Additionally,
sentence\_transformers.util package is used to determine cosine similarity scores between embeddings.

\noindent \textbf{LLM Handler}: openai package is used for automatic communication with ChatGPT models. On the other hand, Gemini does not expose any API endpoint. As a result, we have to provide the prompt to Geimini manually. 

\noindent \textbf{Post-Processor}: json package is used to process the response from LLM. networks package is used to show the graph graphically.

%% file: content/05_result.tex
\section{Results}
We define a system consisting of Oculus, Jetson TX1 and Raspberry Pi as the target of the attack graph generated from vulnerability information.
The system is assumed to be fully connected i.e. each component can communicate with all other components. We use this system to study the performance of LLM under different scenarios.
\subsection{Effect of Context Information in Prompt}
At the beginning, we give the following prompt to ChatGPT which does not have any context information.

\textit{Create an attack graph in json format using only nodes and edges as keys for a system comprising of Raspberry Pi, Oculus Desktop, NVIDIA Jetson Nano. }

\begin{figure}
    \centering
    \includegraphics[width=0.5\textwidth]{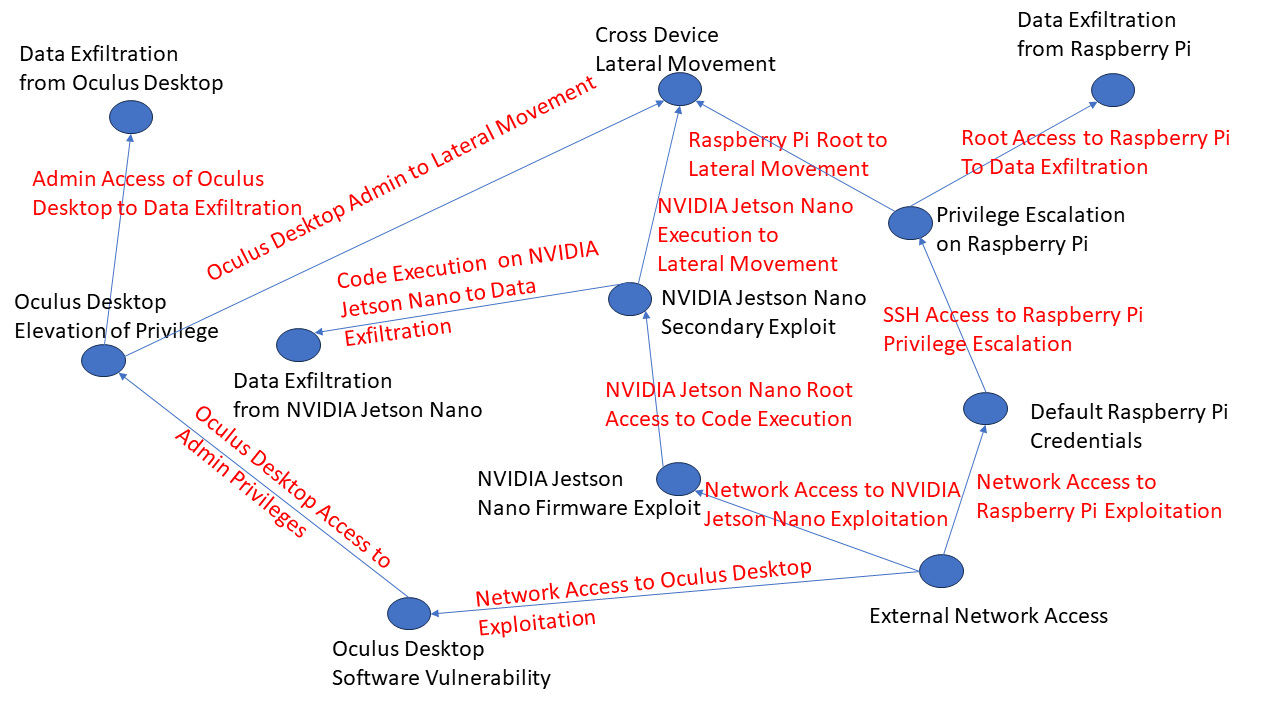}
    \caption{Attack Graph generated by ChatGPT when no context is given in the prompt.}
    \label{fig: without context}
\end{figure}

\begin{figure}
    \centering
    \includegraphics[width=0.5\textwidth]{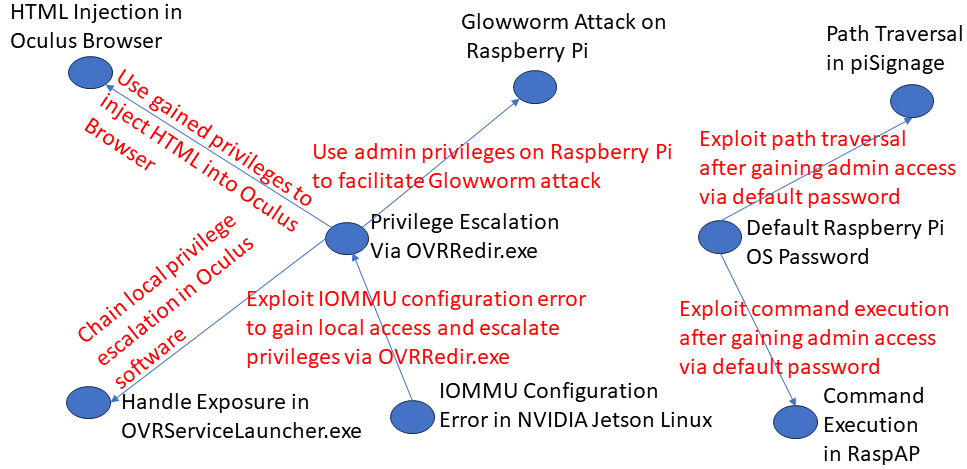}
    \caption{Attack Graph generated by ChatGPT when relevant context is given in the prompt.}
    \label{fig: with context}
\end{figure}

The resulting attack graph is presented in Figure \ref{fig: without context}. This graph is very abstract. Moreover, it does not have any chaining between cross-device vulnerabilities.
After that, we provide vulnerabilities related to the components of the defined system as context. The resulting graph is shown in Figure \ref{fig: with context}. This graph

\subsection{Attack Graph from a Threat Report}
\begin{figure*}
    \centering
    \includegraphics[width=\textwidth]{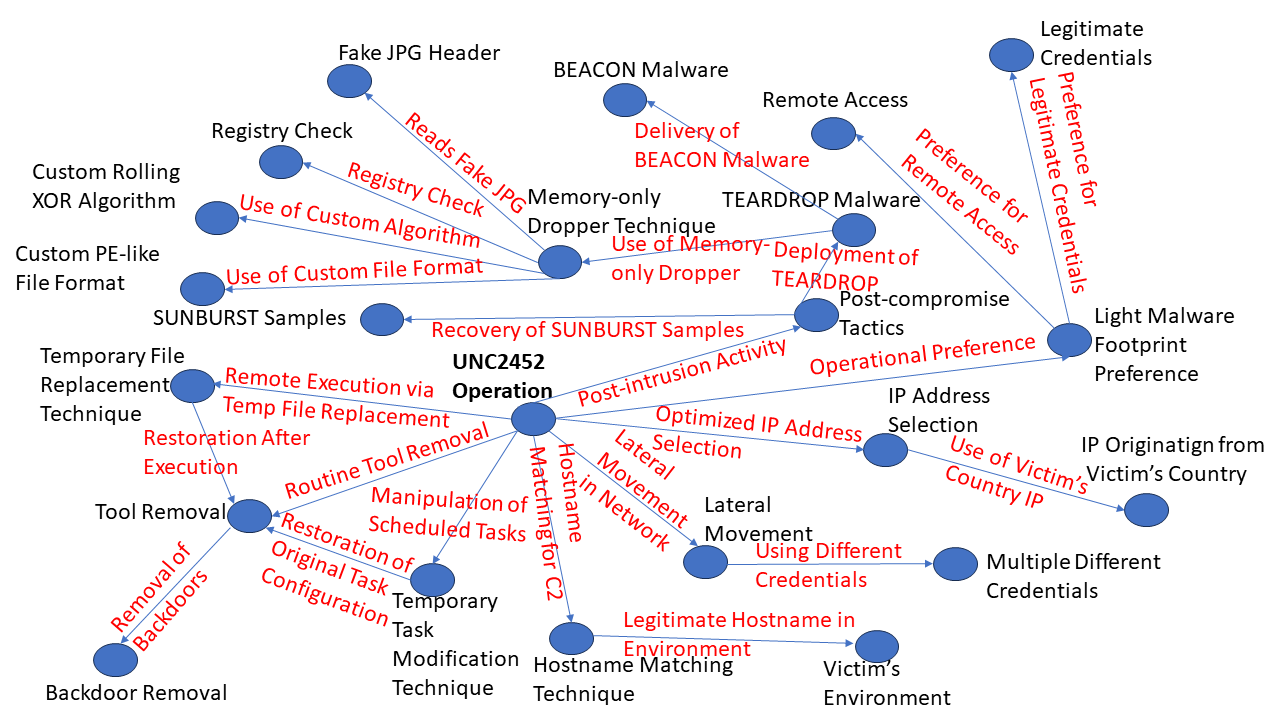}
    \caption{Attack Graph Created from a part of Solar Winds Threat report by GPT-4. This part of the report focuses on the evasion techniques adopted by the attackers}
    \label{fig:threat report}
\end{figure*}

Figure \ref{fig:threat report} shows an attack graph generated from a portion of Solar Winds Incident Threat report by using GPT-4 model. This portion focuses on the evasion tactics that were used by the attackers.The prompt for this attack graph is available in Appendix \ref{appendix: SolarWinds evasion}. On the other hand, Figure \ref{fig: solarwinds attack} shows the attack graph generated from the full threat report. Even though the GPT -4 model is able to process the full incident report, it creates a very high-level attack graph. In Figure \ref{fig: solarwinds attack}, only the node \textit{Operational Security} and its children represent evasion techniques deployed by the attackers despite having access to the same information as when it produces Figure \ref{fig:threat report}. As a result we can conclude that GPT-4 performs better when dealing with smaller prompts.

\subsection{Comparison among Different LLMs}
We evaluate the performances of three different LLM models in generating attack graphs from vulnerability information (CVEs) and from threat reports.

\subsubsection{Attack Graph from Vulnerability Information}
\begin{figure}
    \centering
    \includegraphics[width=0.5\textwidth]{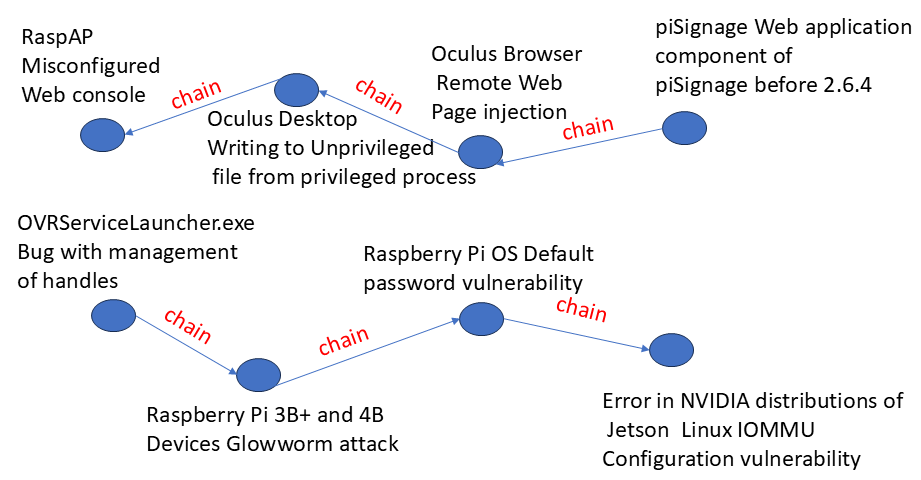}
    \caption{Attack Graph of the defined system generated by GPT-3.5 model}
    \label{fig: system graph gpt 3}
\end{figure}

\begin{figure}
    \centering
    \includegraphics[width=0.5\textwidth]{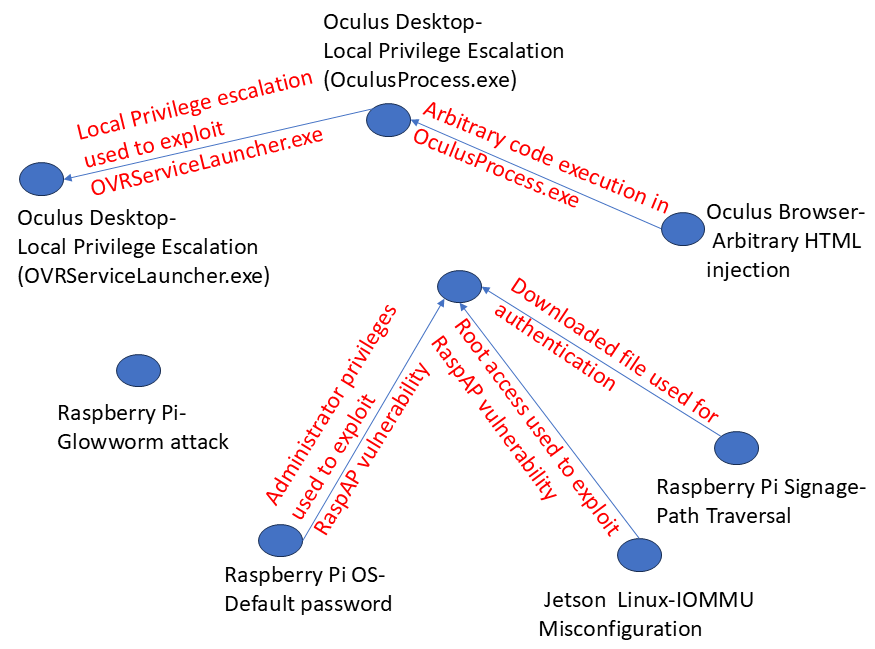}
    \caption{Attack Graph of the defined system generated by Bard model}
    \label{fig: system graph bard}
\end{figure}

Figure \ref{fig: with context}, Figure \ref{fig: system graph gpt 3} and Figure \ref{fig: system graph bard} show attack graphs generated by different LLM models for the system defined at the beginning of this section. Same prompt is sent to each model. The prompt is available in Appendix \ref{appendix: defined system} if interested. 
From these graphs, we can conclude that the performance of the GPT-4 model is the best in terms of the details provided in the graph and cross-device vulnerability chaining.

Figure \ref{fig: with context}, Figure \ref{fig: system graph gpt 3} and Figure \ref{fig: system graph bard} show attack graphs created by different LLM models for the system described at the beginning of this section. The same prompt was sent to each model, and the prompt is available in Appendix \ref{appendix: defined system} if interested. From these graphs, we can observe that the GPT-4 model performs the best in terms of the details provided in the graph and cross-device vulnerability chaining.

\subsubsection{Attack Graph from Threat Report}
\begin{figure}
    \centering
    \includegraphics[width=0.5\textwidth]{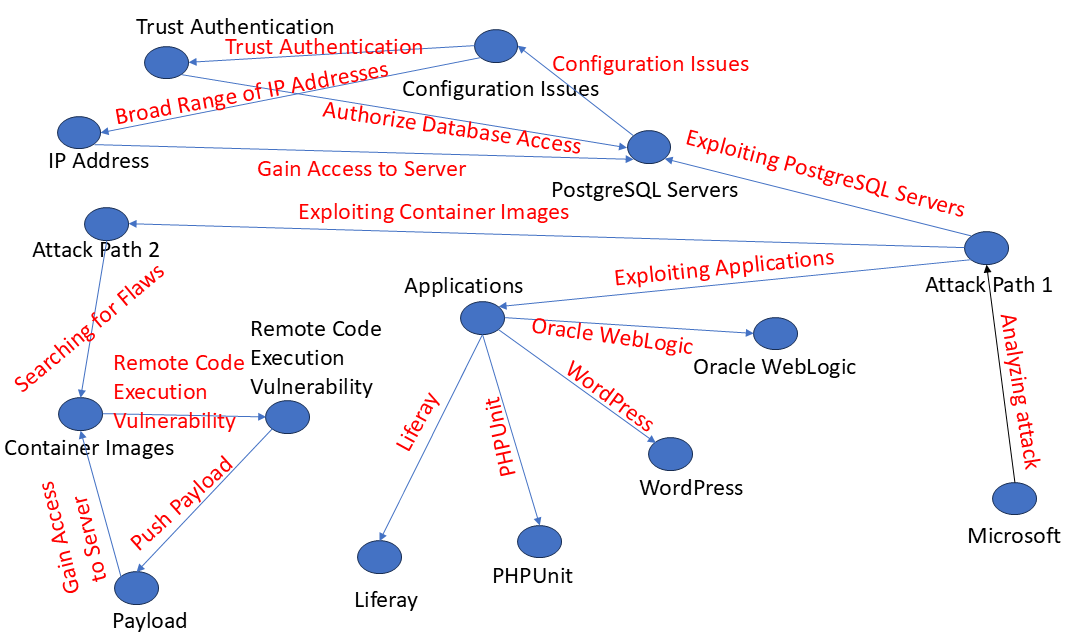}
    \caption{Attack Graph from Kubernetes Cluster Hacked Threat Report generated by GPT-3.5 model}
    \label{fig: kubernetes gpt 3}
\end{figure}

\begin{figure}
    \centering
    \includegraphics[width=0.5\textwidth]{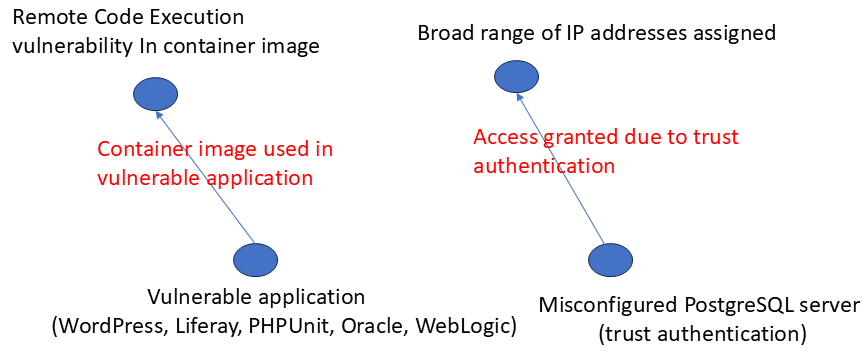}
    \caption{Attack Graph from Kubernetes Cluster Hacked Threat Report generated by BARD model}
    \label{fig: kubernetes bard}
\end{figure}

We use a threat report about Kubernetes clusters being hacked for evaluation. The prompt is available in Appendix \ref{appendix: kubernetes}. In Figure \ref{fig: kubernetes gpt 4}, Figure \ref{fig: kubernetes gpt 3}, and Figure \ref{fig: kubernetes bard}, attack graphs generated by different LLM models are presented. Among these, only the graph produced by the GPT-4 model is able to effectively represent the attack paths in a logically coherent manner. The threat report outlines two distinct attack paths, both of which are accurately captured by the GPT-4 model.

\begin{figure}
    \centering
    \includegraphics[width=0.5\textwidth]{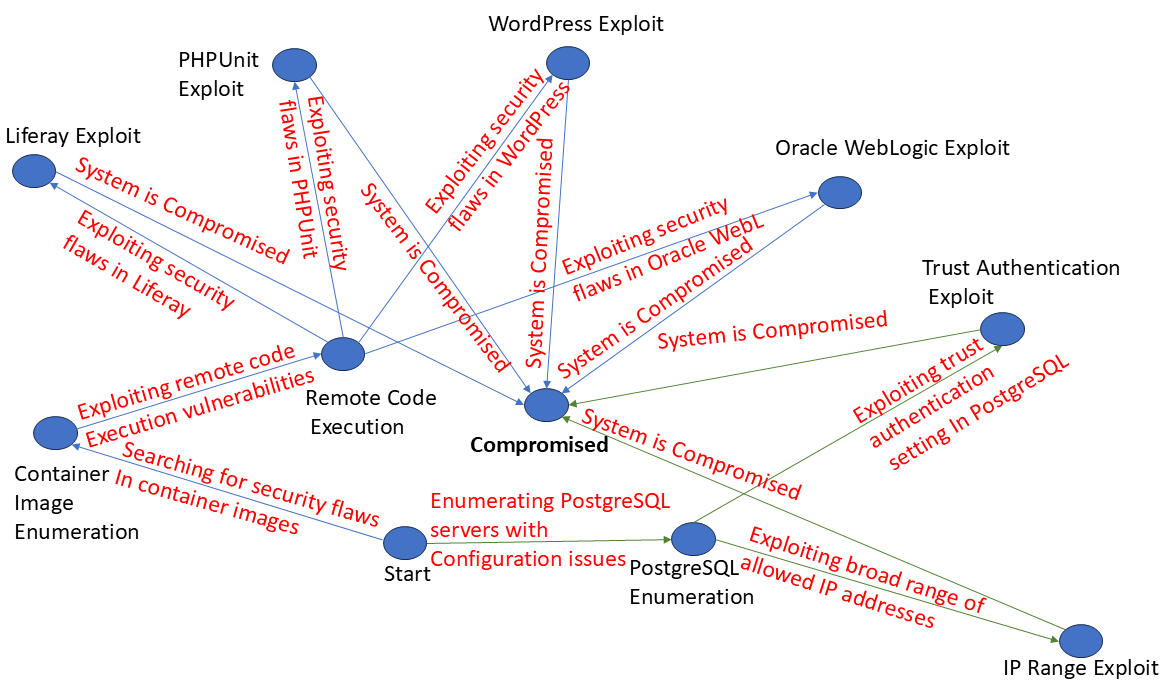}
    \caption{Attack Graph from Kubernetes Cluster Hacked Threat Report generated by GPT-4 model}
    \label{fig: kubernetes gpt 4}
\end{figure}

\subsection{Attempts on Expanding the Attack Graph Gradually}
All popular LLMs have token limits in both prompt and output. As a result, it is hard to get a complete attack graph for a complex system through a single call to LLM. Besides, LLM normally skips low level technical/implementation details and provides high level attack graphs. We explored different ways to address these problems. Our findings are provided below:
\begin{itemize}
    \item We broke down the entire context into smaller chunks based on the token limit of the LLM. Then, we sent multiple requests to the LLM, each with a single chunk as context. After receiving results from each call, we combined them to create the complete graph. To build the full graph, we merged all the nodes and edges from the partial graphs.
    \item Sometimes, answers from LLMs are cut in the middle due to the token limit. We are able to fix the malformed json output and create a partial attack graph from the answer. We also tried to make a follow-up call to the LLM to get the full attack graph. We included the partial answer in the prompt along with the original context and asked LLM to provide the rest of the graph. However, this caused the prompt to exceed the maximum allowed token number.
    \item Attempting to zoom in the attack graph generated by LLM i.e. get more details about the steps depicted in the graph, we made follow up calls to LLM by including an edge from the graph and asking LLM to provide more detailed graph for that particular edge. However, LLM provided the same attack graph again instead of adding more details for the selected edge. Afterwards, we tried another approach where we asked LLM to provide the part of the context which is related to the selected edge. LLM was able to perform that task correctly.
\end{itemize}

%% file: content/06_related_work.tex
\section{Related Work}
Existing works on Attack Graph Generation can be divided into two main approaches. 

One approach \cite{jha2002two,noel2002combinatorial,ou2005mulval,al2019a2g2v,sheyner2002automated,ou2006scalable,hankin2022attack} uses model reasoning/combinatorial analysis to create attack graphs. \cite{ou2005mulval} presents MulVAL, an end-to-end framework and reasoning system for vulnerability analysis on a network. MulVAL uses Datalog as the modeling language.  The information in the vulnerability database are encoded as Datalog facts. The reasoning engine consists of a collection of Datalog rules. These rules describe the behavior of the operating system and how the different components in the network interact with each other. \cite{al2019a2g2v} proposes
a model-checking-based automated attack graph generator
and visualizer called A2G2V. A2G2V tool uses system architecture and atomic attack behaviors captured in AADL and AGREE Annex as input. The AADL+AGREE model is translated into its Lustre
equivalent, and further analyzed by using the model-checker
JKind. A2G2V tool uses JKind model-checker to generate attack sequences by iteratively relaxing specifications whereas Graphviz tool is integrated for attack graph visualization. \cite{noel2002combinatorial} establishes encoding rules to reason about interdependent vulnerabilities and exploits. The combinatorial analysis of network vulnerability models attacker exploits in terms of exploit pre- and post-conditions to capture the interdependencies of the exploits. Inference engines then reason about combinations of exploits with the goal of discovering attack paths for the assumed attacker goals.  Model reasoning and combinatorial analysis based approaches are highly accurate and provide a deep understanding of potential attack paths. However, they are computationally intensive and do not scale well with large, complex networks. Moreover, these works require manual input of related vulnerability information for a target system. 

Another approach \cite{payne2019secure,bezawada2019agbuilder,jin2023prometheus,husari2017ttpdrill} for automatic attack graph generation involves using natural language processing (NLP) techniques and machine learning models. \cite{payne2019secure} introduces the concept of attack circuits, which are created using input/output pairs derived from CVEs with the aid of NLP, and an attack graph that is composed of these circuits. To construct the attack circuits, vulnerability descriptions from a vulnerability database are used. For each item in the database, an input/output pair is generated, where the input represents the attack source and the output represents the attack target. \cite{bezawada2019agbuilder} models the attack graph generation and analysis problem as a planning problem. For that purpose, the attack graph is encoded in the Planner Domain Definition Language (PDDL) representation, referred to as a PDDL domain. The corresponding tool, AGBuilder –Attack Graph Builder, is designed to automatically generate PDDL based representation of attacks from textual description of vulnerabilities found in the CVE system or the NVD system. \cite{husari2017ttpdrill} develops a tool called TTPDrill which performs automated and context-aware analysis of cyber threat intelligence (CTI) and learns attack patterns (TTPs) from commonly available CTI sources. TTPDrill uses a novel text mining approach extracts threat actions based on semantic relationship. In addition, \cite{husari2017ttpdrill} presents a novel threat-action ontology that is sufficiently rich to understand the specifications and context of malicious actions. The NLP and machine learning based approach can efficiently process large volumes of data and adapt to new threats. However, training these machine learning models require significant amount of quality training data, time and computational resources. 

%% file: content/07_conclusion.tex
\section{Conclusion}
In this paper, we have delved into the novel application of large language models like ChatGPT for the task of generating attack graphs. Our focus has been on the automatic chaining of Common Vulnerabilities and Exposures (CVEs) based on their preconditions and postconditions as well as generating attack graphs from threat reports. The findings suggest that ChatGPT not only effectively generates but also enhances the quality of attack graphs by introducing greater contextual relevance and nuanced understanding.

However, it is important to note the limitations of this approach. While the language model performs well in many cases, the lack of domain-specific expertise can sometimes result in graphs that may need further refinement or validation. Moreover, the ethical implications of using machine learning models for cybersecurity tasks, including the potential for misuse, require thorough scrutiny.

In summary, the application of large language models like ChatGPT in the cybersecurity domain, specifically for generating attack graphs, reveals promising results that merit deeper investigation and broader application. The potential benefits in terms of efficiency, accuracy, and real-time relevance make it a compelling approach for modern-day security challenges.

%% file: content/08_appendix.tex
\appendix
\section{Prompt for Attack Graph of Defined System} 
\label{appendix: defined system}
Create an attack graph in json format using only nodes and edges as keys from the vulnerability information given below. node should have id, label, precondition and postcondition as properties. edge should have from, to and label properties. Do not give a simplified graph. Add as much detail as possible. Incorporate all possible information in nodes and edges. Do not create separate keys for them in the json. Chain the vulnerabilities if applicable. Vulnerabilities can be chained if precondition of one vulnerability is similar to the post condition of another vulnerability. Vulnerability with matching postcondition should be ahead in the chain. Return only the json as response. Do not put any text before or after the json.

The web application component of piSignage before 2.6.4 allows a remote attacker (authenticated as a low-privilege user) to download arbitrary files from the Raspberry Pi via api/settings/log?file=../ path traversal. In other words, this issue is in the player API for log download.

A remote web page could inject arbitrary HTML code into the Oculus Browser UI, allowing an attacker to spoof UI and potentially execute code. This affects the Oculus Browser starting from version 5.2.7 until 5.7.11.

Writing to an unprivileged file from a privileged OVRRedir.exe process in Oculus Desktop before 1.44.0.32849 on Windows allows local users to write to arbitrary files and consequently gain privileges via vectors involving a hard link to a log file.

An issue was discovered in includes/webconsole.php in RaspAP 2.5. With authenticated access, an attacker can use a misconfigured (and virtually unrestricted) web console to attack the underlying OS (Raspberry Pi) running this software, and execute commands on the system (including ones for uploading of files and execution of code).

Due to a bug with management of handles in OVRServiceLauncher.exe, an attacker could expose a privileged process handle to an unprivileged process, leading to local privilege escalation. This issue affects Oculus Desktop versions after 1.39 and prior to 31.1.0.67.507.

Raspberry Pi 3 B+ and 4 B devices through 2021-08-09, in certain specific use cases in which the device supplies power to audio-output equipment, allow remote attackers to recover speech signals from an LED on the device, via a telescope and an electro-optical sensor, aka a "Glowworm" attack. We assume that the Raspberry Pi supplies power to some speakers. The power indicator LED of the Raspberry Pi is connected directly to the power line, as a result, the intensity of a device\'s power indicator LED is correlative to the power consumption. The sound played by the speakers affects the Raspberry Pi\'s power consumption and as a result is also correlative to the light intensity of the LED. By analyzing measurements obtained from an electro-optical sensor directed at the power indicator LED of the Raspberry Pi, we can recover the sound played by the speakers.

Raspberry Pi OS through 5.10 has the raspberry default password for the pi account. If not changed, attackers can gain administrator privileges.

NVIDIA distributions of Jetson Linux contain a vulnerability where an error in the IOMMU configuration may allow an unprivileged attacker with physical access to the board direct read/write access to the entire system address space through the PCI bus. Such an attack could result in denial of service, code execution, escalation of privileges, and impact to data integrity and confidentiality. The scope impact may extend to other components.

\section{Prompt for Attack Graph from Kubernetes Cluster Hacked Threat Report}
\label{appendix: kubernetes}
Create an attack graph in json format using nodes and edges from following attack scenario: The security researchers at Microsoft analyzed the attack and identified two attack paths were used. The first attack path is establishing and enumerating the PostgreSQL servers that had configuration issues. From there one of the most common misconfigurations that were being exploited is the “trust authentication” setting which allows PostgreSQL to make an assumption that any connection that is established towards the server is authorized to get database access. In addition, if a security issue exists such that a broad range of IP addresses are being assigned then any IP address that the attacker may be using can be used to gain access to the server. The second attack path is trying to exploit a security flaw in container images. In this particular scenario, the attackers are searching for a remote code execution vulnerability which will then allow them to push their payload and gain access to the server in that manner. From what has been seen so far, the attackers are trying to find and exploit security flaws in these applications: WordPress Liferay PHPUnit Oracle WebLogic.

\section{Prompt for Attack Graph from SolarWinds Attack Threat Report}
\subsection{Full Report}
\label{appendix: SolarWinds full}
Create an attack graph in json format using only nodes and edges as keys from the attack scenario given below. node should have id, label, edge should have from, to and label properties. Do not give a simplified graph. Add as much detail as possible. Incorporate all possible information in nodes and edges. Do not create separate keys for them in the json. Give only the graph in json format. Do not put text before or after json.

Executive Summary
We have discovered a global intrusion campaign. We are tracking the actors behind this campaign as UNC2452.
FireEye discovered a supply chain attack trojanizing SolarWinds Orion business software updates in order to distribute malware we call SUNBURST. 
The attacker’s post compromise activity leverages multiple techniques to evade detection and obscure their activity, but these efforts also offer some opportunities for detection.
The campaign is widespread, affecting public and private organizations around the world.
FireEye is releasing signatures to detect this threat actor and supply chain attack in the wild. These are found on our public GitHub page. FireEye products and services can help customers detect and block this attack.
Summary
FireEye has uncovered a widespread campaign, that we are tracking as UNC2452. The actors behind this campaign gained access to numerous public and private organizations around the world. They gained access to victims via trojanized updates to SolarWind’s Orion IT monitoring and management software. This campaign may have begun as early as Spring 2020 and is currently ongoing. Post compromise activity following this supply chain compromise has included lateral movement and data theft. The campaign is the work of a highly skilled actor and the operation was conducted with significant operational security.

SUNBURST Backdoor
SolarWinds.Orion.Core.BusinessLayer.dll is a SolarWinds digitally-signed component of the Orion software framework that contains a backdoor that communicates via HTTP to third party servers. We are tracking the trojanized version of this SolarWinds Orion plug-in as SUNBURST.

After an initial dormant period of up to two weeks, it retrieves and executes commands, called “Jobs”, that include the ability to transfer files, execute files, profile the system, reboot the machine, and disable system services. The malware masquerades its network traffic as the Orion Improvement Program (OIP) protocol and stores reconnaissance results within legitimate plugin configuration files allowing it to blend in with legitimate SolarWinds activity. The backdoor uses multiple obfuscated blocklists to identify forensic and anti-virus tools running as processes, services, and drivers.
Multiple trojanzied updates were digitally signed from March - May 2020 and posted to the SolarWinds updates website, including:

hxxps://downloads.solarwinds[.]com/solarwinds/CatalogResources/
Core/2019.4/2019.4.5220.20574/SolarWinds-Core-v2019.4.5220-Hotfix5.msp
The trojanized update file is a standard Windows Installer Patch file that includes compressed resources associated with the update, including the trojanized SolarWinds.Orion.Core.BusinessLayer.dll component. Once the update is installed, the malicious DLL will be loaded by the legitimate SolarWinds.BusinessLayerHost.exe or SolarWinds.BusinessLayerHostx64.exe (depending on system configuration). After a dormant period of up to two weeks, the malware will attempt to resolve a subdomain of avsvmcloud[.]com. The DNS response will return a CNAME record that points to a Command and Control (C2) domain. The C2 traffic to the malicious domains is designed to mimic normal SolarWinds API communications. The list of known malicious infrastructure is available on FireEye’s GitHub page.

Worldwide Victims Across Multiple Verticals
FireEye has detected this activity at multiple entities worldwide. The victims have included government, consulting, technology, telecom and extractive entities in North America, Europe, Asia and the Middle East. We anticipate there are additional victims in other countries and verticals. FireEye has notified all entities we are aware of being affected.

Post Compromise Activity and Detection Opportunities
We are currently tracking the software supply chain compromise and related post intrusion activity as UNC2452. After gaining initial access, this group uses a variety of techniques to disguise their operations while they move laterally (Figure 2). This actor prefers to maintain a light malware footprint, instead preferring legitimate credentials and remote access for access into a victim’s environment.

This section will detail the notable techniques and outline potential opportunities for detection.

TEARDROP and BEACON Malware Used

Multiple SUNBURST samples have been recovered, delivering different payloads. In at least one instance the attackers deployed a previously unseen memory-only dropper we’ve dubbed TEARDROP to deploy Cobalt Strike BEACON.

TEARDROP is a memory only dropper that runs as a service, spawns a thread and reads from the file “gracious\_truth.jpg”, which likely has a fake JPG header. Next it checks that HKU\\SOFTWARE\\Microsoft\\CTF exists, decodes an embedded payload using a custom rolling XOR algorithm and manually loads into memory an embedded payload using a custom PE-like file format. TEARDROP does not have code overlap with any previously seen malware. We believe that this was used to execute a customized Cobalt Strike BEACON.

Mitigation: FireEye has provided two Yara rules to detect TEARDROP available on our GitHub. Defenders should look for the following alerts from FireEye HX: MalwareGuard and WindowsDefender:

Process Information

file\_operation\_closed
file-path*: “c:\\windows\\syswow64\\netsetupsvc.dll
actor-process:
pid: 17900

Window’s defender Exploit Guard log entries: (Microsoft-Windows-Security-Mitigations/KernelMode event ID 12)           

Process”\\Device\\HarddiskVolume2\\Windows\\System32\\svchost.exe” (PID XXXXX) would have been blocked from loading the non-Microsoft-signed binary
‘\\Windows\\SysWOW64\\NetSetupSvc.dll’

Attacker Hostnames Match Victim Environment

The actor sets the hostnames on their command and control infrastructure to match a legitimate hostname found within the victim’s environment. This allows the adversary to blend into the environment, avoid suspicion, and evade detection.

Detection Opportunity

The attacker infrastructure leaks its configured hostname in RDP SSL certificates, which is identifiable in internet-wide scan data. This presents a detection opportunity for defenders -- querying internet-wide scan data sources for an organization’s hostnames can uncover malicious IP addresses that may be masquerading as the organization. (Note: IP Scan history often shows IPs switching between default (WIN-*) hostnames and victim’s hostnames) Cross-referencing the list of IPs identified in internet scan data with remote access logs may identify evidence of this actor in an environment. There is likely to be a single account per IP address.

IP Addresses located in Victim’s Country

The attacker’s choice of IP addresses was also optimized to evade detection. The attacker primarily used only IP addresses originating from the same country as the victim, leveraging Virtual Private Servers.

Detection Opportunity

This also presents some detection opportunities, as geolocating IP addresses used for remote access may show an impossible rate of travel if a compromised account is being used by the legitimate user and the attacker from disparate IP addresses. The attacker used multiple IP addresses per VPS provider, so once a malicious login from an unusual ASN is identified, looking at all logins from that ASN can help detect additional malicious activity. This can be done alongside baselining and normalization of ASN’s used for legitimate remote access to help identify suspicious activity.

Lateral Movement Using Different Credentials

Once the attacker gained access to the network with compromised credentials, they moved laterally using multiple different credentials. The credentials used for lateral movement were always different from those used for remote access.

Detection Opportunity

Organizations can use HX’s LogonTracker module to graph all logon activity and analyze systems displaying a one-to-many relationship between source systems and accounts. This will uncover any single system authenticating to multiple systems with multiple accounts, a relatively uncommon occurrence during normal business operations.

Temporary File Replacement and Temporary Task Modification

The attacker used a temporary file replacement technique to remotely execute utilities: they replaced a legitimate utility with theirs, executed their payload, and then restored the legitimate original file. They similarly manipulated scheduled tasks by updating an existing legitimate task to execute their tools and then returning the scheduled task to its original configuration. They routinely removed their tools, including removing backdoors once legitimate remote access was achieved.

Detection Opportunity

Defenders can examine logs for SMB sessions that show access to legitimate directories and follow a delete-create-execute-delete-create pattern in a short amount of time. Additionally, defenders can monitor existing scheduled tasks for temporary updates, using frequency analysis to identify anomalous modification of tasks. Tasks can also be monitored to watch for legitimate Windows tasks executing new or unknown binaries.

This campaign’s post compromise activity was conducted with a high regard for operational security, in many cases leveraging dedicated infrastructure per intrusion. This is some of the best operational security that FireEye has observed in a cyber attack, focusing on evasion and leveraging inherent trust. However, it can be detected through persistent defense.

In-Depth Malware Analysis
SolarWinds.Orion.Core.BusinessLayer.dll (b91ce2fa41029f6955bff20079468448) is a SolarWinds-signed plugin component of the Orion software framework that contains an obfuscated backdoor which communicates via HTTP to third party servers. After an initial dormant period of up to two weeks, it retrieves and executes commands, called “Jobs”, that include the ability to transfer and execute files, profile the system, and disable system services. The backdoor’s behavior and network protocol blend in with legitimate SolarWinds activity, such as by masquerading as the Orion Improvement Program (OIP) protocol and storing reconnaissance results within plugin configuration files. The backdoor uses multiple blocklists to identify forensic and anti-virus tools via processes, services, and drivers.

Unique Capabilities
Subdomain DomainName Generation Algorithm (DGA) is performed to vary DNS requests
CNAME responses point to the C2 domain for the malware to connect to
The IP block of A record responses controls malware behavior
DGA encoded machine domain name, used to selectively target victims
Command and control traffic masquerades as the legitimate Orion Improvement Program
Code hides in plain site by using fake variable names and tying into legitimate components
Delivery and Installation
Authorized system administrators fetch and install updates to SolarWinds Orion via packages distributed by SolarWinds’s website. The update package CORE-2019.4.5220.20574-SolarWinds-Core-v2019.4.5220-Hotfix5.msp (02af7cec58b9a5da1c542b5a32151ba1) contains the SolarWinds.Orion.Core.BusinessLayer.dll described in this report. After installation, the Orion software framework executes the .NET program SolarWinds.BusinessLayerHost.exe to load plugins, including SolarWinds.Orion.Core.BusinessLayer.dll. This plugin contains many legitimate namespaces, classes, and routines that implement functionality within the Orion framework. Hidden in plain sight, the class SolarWinds.Orion.Core.BusinessLayer.
OrionImprovementBusinessLayer implements an HTTP-based backdoor. Code within the logically unrelated routine SolarWinds.Orion.Core.
BusinessLayer.BackgroundInventory.InventoryManager.RefreshInternal invokes the backdoor code when the Inventory Manager plugin is loaded.

SolarWinds.Orion.Core.BusinessLayer.dll is signed by SolarWinds, using the certificate with serial number 0f:e9:73:75:20:22:a6:06:ad:f2:a3:6e:
34:5d:c0:ed. The file was signed on March 24, 2020.

Initialization

On execution of the malicious SolarWinds.Orion.Core.BusinessLayer.
OrionImprovementBusinessLayer.Initialize method the sample verifies that its lower case process name hashes to the value \\
17291806236368054941. This hash value is calculated as the standard FNV-1A 64-bit hash with an additional XOR by 6605813339339102567 after computing the FNV-1A. This hash matches a process named "solarwinds.businesslayerhost".

The sample only executes if the filesystem write time of the assembly is at least 12 to 14 days prior to the current time; the exact threshold is selected randomly from an interval. The sample continues to check this time threshold as it is run by a legitimate recurring background task. Once the threshold is met, the sample creates the named pipe 583da945-62af-10e8-4902-a8f205c72b2e to act as a guard that only one instance is running before reading SolarWinds.Orion.Core.BusinessLayer.dll.config from disk and retrieving the XML field appSettings. The appSettings fields’ keys are legitimate values that the malicious logic re-purposes as a persistent configuration. The key ReportWatcherRetry must be any value other than 3 for the sample to continue execution.

The sample checks that the machine is domain joined and retrieves the domain name before execution continues. A userID is generated by computing the MD5 of a network interface MAC address that is up and not a loopback device, the domain name, and the registry value HKEY\_LOCAL\_MACHINE\\SOFTWARE\\Microsoft\\Cryptography\\MachineGuid. The userID is encoded via a custom XOR scheme after the MD5 is calculated. The ReportWatcherPostpone key of appSettings is then read from SolarWinds.Orion.Core.BusinessLayer.dll.config to retrieve the initial, legitimate value. This operation is performed as the sample later bit packs flags into this field and the initial value must be known in order to read out the bit flags. The sample then invokes the method Update which is the core event loop of the sample.

DGA and Blocklists
The backdoor determines its C2 server using a Domain Generation Algorithm (DGA) to construct and resolve a subdomain of avsvmcloud[.]com. The Update method is responsible for initializing cryptographic helpers for the generation of these random C2 subdomains. Subdomains are generated by concatenating a victim userId with a reversible encoding of the victims local machine domain name. The attacker likely utilizes the DGA subdomain to vary the DNS response to victims as a means to control the targeting of the malware. These subdomains are concatenated with one of the following to create the hostname to resolve:

.appsync-api.eu-west-1[.]avsvmcloud[.]com
.appsync-api.us-west-2[.]avsvmcloud[.]com
.appsync-api.us-east-1[.]avsvmcloud[.]com
.appsync-api.us-east-2[.]avsvmcloud[.]com
Process name, service name, and driver path listings are obtained, and each value is hashed via the FNV-1a + XOR algorithm as described previously and checked against hardcoded blocklists. Some of these hashes have been brute force reversed as part of this analysis, showing that these routines are scanning for analysis tools and antivirus engine components. If a blocklisted process is found the Update routine exits and the sample will continue to try executing the routine until the blocklist passes. Blocklisted services are stopped by setting their HKLM\\SYSTEM\\CurrentControlSet\\services\\<service\_name>\\Start registry entries to value 4 for disabled. Some entries in the service list if found on the system may affect the DGA algorithms behavior in terms of the values generated. The list of stopped services is then bit-packed into the ReportWatcherPostpone key of the appSettings entry for the samples’ config file. If any service was transitioned to disabled the Update method exits and retries later. The sample retrieves a driver listing via the WMI query Select * From Win32\_SystemDriver. If any blocklisted driver is seen the Update method exits and retries. If all blocklist tests pass, the sample tries to resolve api.solarwinds.com to test the network for connectivity.

Network Command and Control (C2)
If all blocklist and connectivity checks pass, the sample starts generating domains in a while loop via its DGA. The sample will delay for random intervals between the generation of domains; this interval may be any random value from the ranges 1 to 3 minutes, 30 to 120 minutes, or on error conditions up to 420 to 540 minutes (9 hours). The DNS A record of generated domains is checked against a hardcoded list of IP address blocks which control the malware’s behavior. Records within the following ranges will terminate the malware and update the configuration key ReportWatcherRetry to a value that prevents further execution:

10.0.0.0/8
172.16.0.0/12
192.168.0.0/16
224.0.0.0/3
fc00:: - fe00::
fec0:: - ffc0::
ff00:: - ff00::
20.140.0.0/15
96.31.172.0/24
131.228.12.0/22
144.86.226.0/24
Once a domain has been successfully retrieved in a CNAME DNS response the sample will spawn a new thread of execution invoking the method HttpHelper.Initialize which is responsible for all C2 communications and dispatching. The HTTP thread begins by delaying for a configurable amount of time that is controlled by the SetTime command. The HTTP thread will delay for a minimum of 1 minute between callouts. The malware uses HTTP GET or HTTP POST requests. If the sample is attempting to send outbound data the content-type HTTP header will be set to "application/octet-stream" otherwise to "application/json".

A JSON payload is present for all HTTP POST and PUT requests and contains the keys “userId”, “sessionId”, and “steps”. The “steps” field contains a list of objects with the following keys: “Timestamp”, “Index”, “EventType”, “EventName”, “DurationMs”, “Succeeded”, and “Message”. The JSON key “EventType” is hardcoded to the value “Orion”, and the “EventName” is hardcoded to “EventManager”. Malware response messages to send to the server are DEFLATE compressed and single-byte-XOR encoded, then split among the “Message” fields in the “steps” array. Each “Message” value is Base64 encoded separately. Not all objects in the “steps” array contribute to the malware message – the integer in the “Timestamp” field must have the 0x2 bit set to indicate that the contents of the “Message” field are used in the malware message. Step objects whose bit 0x2 is clear in the Timestamp field contain random data and are discarded when assembling the malware response.

Steganography
In observed traffic these HTTP response bodies attempt to appear like benign XML related to .NET assemblies, but command data is actually spread across the many GUID and HEX strings present. Commands are extracted from HTTP response bodies by searching for HEX strings using the following regular expression: "\{[0-9a-f-]{36}\}"|"[0-9a-f]{32}"|"[0-9a-f]{16}". Command data is spread across multiple strings that are disguised as GUID and HEX strings. All matched substrings in the response are filtered for non HEX characters, joined together, and HEX-decoded. The first DWORD value shows the actual size of the message, followed immediately with the message, with optional additional junk bytes following. The extracted message is single-byte XOR decoded using the first byte of the message, and this is then DEFLATE decompressed. The first character is an ASCII integer that maps to the JobEngine enum, with optional additional command arguments delimited by space characters.

Commands are then dispatched to a JobExecutionEngine based upon the command value as described next.

Supported Commands
Command	Value	Operation
Idle	0	No operation
Exit	1	Terminate the current thread.
SetTime	2	Sets the delay time between main event loop executions Delay is in seconds, and varies random between [.9 * <delay>, 1.1 * <delay>].          If the delay is < 300 it is doubled on the next execution through the loop, this means it should settle onto an interval of around [5, 10] minutes.         There is a second, unrelated delay routine that delays for a random interval between [16hrs, 83hrs]
CollectSystemDescription	3	Profile the local system including hostname, username, OS version, MAC addresses, IP address, DHCP configuration, and domain information.
UploadSystemDescription	4	Perform a HTTP request to the specified URL, parse the results and compare components against unknown hashed values. Format a report and send to the C2 server.
RunTask	5	Starts a new process with the given file path and arguments
GetProcessByDescription	6	
Returns a process listing. If no arguments are provided returns just the PID and process name.        If an argument is provided it also returns the parent PID and username and domain for the process owner.

KillTask	7	Terminate the given process, by PID.
GetFileSystemEntries	8	Given a path and an optional match pattern recursively list files and directories
WriteFile	9	Given a file path and a Base64 encoded string write the contents of the Base64 decoded string to the given file path. Write using append mode. Delay for [1s, 2s] after writing is done.
FileExists	10	Tests whether the given file path exists.
DeleteFile	11	Deletes the specified file path.
GetFileHash	12	
Compute the MD5 of a file at a given path and return result as a HEX string. If an argument is provided, it is the expected MD5 hash of the file and returns an error if the calculated MD5 differs.

ReadRegistryValue	13	Arbitrary registry read from one of the supported hives
SetRegistryValue	14	Arbitrary registry write from one of the supported hives.
DeleteRegistryValue	15	Arbitrary registry delete from one of the supported hives
GetRegistrySubKeyAndValueNames

16	Returns listing of subkeys and value names beneath the given registry path
Reboot	17	Attempts to immediately trigger a system reboot.
Indicators and Detections to Help the Community
To empower the community to detect this supply chain backdoor, we are publishing indicators and detections to help organizations identify this backdoor and this threat actor. The signatures are a mix of Yara, IOC, and Snort formats.

A list of the detections and signatures are available on the Mandiant GitHub repository. We are releasing detections and will continue to update the public repository with overlapping detections for host and network-based indicators as we develop new or refine existing ones. We have found multiple hashes with this backdoor and we will post updates of those hashes.

MITRE ATT\&CK Techniques Observed
ID	Description
T1012	Query Registry
T1027	Obfuscated Files or Information
T1057	Process Discovery
T1070.004	File Deletion
T1071.001	Web Protocols
T1071.004	Application Layer Protocol: DNS
T1083	File and Directory Discovery
T1105	Ingress Tool Transfer
T1132.001	Standard Encoding
T1195.002	Compromise Software Supply Chain
T1518	Software Discovery
T1518.001	Security Software Discovery
T1543.003	Windows Service
T1553.002	Code Signing
T1568.002	Domain Generation Algorithms
T1569.002	Service Execution
T1584	Compromise Infrastructure
Immediate Mitigation Recommendations
Prior to following SolarWind’s recommendation to utilize Orion Platform release 2020.2.1 HF 1, which is currently available via the SolarWinds Customer Portal, organizations should consider preserving impacted devices and building new systems using the latest versions. Applying an upgrade to an impacted box could potentially overwrite forensic evidence as well as leave any additional backdoors on the system. In addition, SolarWinds has released additional mitigation and hardening instructions.

In the event you are unable to follow SolarWinds’ recommendations, the following are immediate mitigation techniques that could be deployed as first steps to address the risk of trojanized SolarWinds software in an environment. If attacker activity is discovered in an environment, we recommend conducting a comprehensive investigation and designing and executing a remediation strategy driven by the investigative findings and details of the impacted environment.

Ensure that SolarWinds servers are isolated / contained until a further review and investigation is conducted. This should include blocking all Internet egress from SolarWinds servers.
If SolarWinds infrastructure is not isolated, consider taking the following steps:
Restrict scope of connectivity to endpoints from SolarWinds servers, especially those that would be considered Tier 0 / crown jewel assets
Restrict the scope of accounts that have local administrator privileged on SolarWinds servers.
Block Internet egress from servers or other endpoints with SolarWinds software.
Consider (at a minimum) changing passwords for accounts that have access to SolarWinds servers / infrastructure. Based upon further review / investigation, additional remediation measures may be required.
If SolarWinds is used to managed networking infrastructure, consider conducting a review of network device configurations for unexpected / unauthorized modifications. Note, this is a proactive measure due to the scope of SolarWinds functionality, not based on investigative findings.

\subsection{Evasion Techniques}
\label{appendix: SolarWinds evasion}
Create an attack graph in json format using only nodes and edges as keys from the attack scenario given below. node should have id, label, edge should have from, to and label properties. Do not give a simplified graph. Add as much detail as possible. Incorporate all possible information in nodes and edges. Do not create separate keys for them in the json. Give only the graph in json format. Do not put text before or after json.

We are currently tracking the software supply chain compromise and related post intrusion activity as UNC2452. After gaining initial access, this group uses a variety of techniques to disguise their operations while they move laterally (Figure 2). This actor prefers to maintain a light malware footprint, instead preferring legitimate credentials and remote access for access into a victim’s environment. Post-compromise tactics TEARDROP and BEACON Malware Used Multiple SUNBURST samples have been recovered, delivering different payloads. In at least one instance the attackers deployed a previously unseen memory-only dropper we’ve dubbed TEARDROP to deploy Cobalt Strike BEACON. TEARDROP is a memory only dropper that runs as a service, spawns a thread and reads from the file “gracious\_truth.jpg”, which likely has a fake JPG header. Next it checks that HKU\\SOFTWARE\\Microsoft\\CTF exists, decodes an embedded payload using a custom rolling XOR algorithm and manually loads into memory an embedded payload using a custom PE-like file format. TEARDROP does not have code overlap with any previously seen malware. We believe that this was used to execute a customized Cobalt Strike BEACON. Attacker Hostnames Match Victim Environment The actor sets the hostnames on their command and control infrastructure to match a legitimate hostname found within the victim’s environment. This allows the adversary to blend into the environment, avoid suspicion, and evade detection. IP Addresses located in Victim’s Country The attacker’s choice of IP addresses was also optimized to evade detection. The attacker primarily used only IP addresses originating from the same country as the victim, leveraging Virtual Private Servers. Lateral Movement Using Different Credentials Once the attacker gained access to the network with compromised credentials, they moved laterally using multiple different credentials. The credentials used for lateral movement were always different from those used for remote access. Temporary File Replacement and Temporary Task Modification The attacker used a temporary file replacement technique to remotely execute utilities: they replaced a legitimate utility with theirs, executed their payload, and then restored the legitimate original file. They similarly manipulated scheduled tasks by updating an existing legitimate task to execute their tools and then returning the scheduled task to its original configuration. They routinely removed their tools, including removing backdoors once legitimate remote access was achieved.

%% file: main.bbl
\begin{thebibliography}{10}

\bibitem{al2019a2g2v}
Alaa~T Al~Ghazo, Mariam Ibrahim, Hao Ren, and Ratnesh Kumar.
\newblock A2g2v: Automatic attack graph generation and visualization and its
  applications to computer and scada networks.
\newblock {\em IEEE Transactions on Systems, Man, and Cybernetics: Systems},
  50(10):3488--3498, 2019.

\bibitem{sheyner2002automated}
Oleg Sheyner, Joshua Haines, Somesh Jha, Richard Lippmann, and Jeannette~M
  Wing.
\newblock Automated generation and analysis of attack graphs.
\newblock In {\em Proceedings 2002 IEEE Symposium on Security and Privacy},
  pages 273--284. IEEE, 2002.

\bibitem{ou2006scalable}
Xinming Ou, Wayne~F Boyer, and Miles~A McQueen.
\newblock A scalable approach to attack graph generation.
\newblock In {\em Proceedings of the 13th ACM conference on Computer and
  communications security}, pages 336--345, 2006.

\bibitem{payne2019secure}
Josh Payne, Karan Budhraja, and Ashish Kundu.
\newblock How secure is your iot network?
\newblock In {\em 2019 IEEE International Congress on Internet of Things
  (ICIOT)}, pages 181--188. IEEE, 2019.

\bibitem{bezawada2019agbuilder}
Bruhadeshwar Bezawada, Indrajit Ray, and Kushagra Tiwary.
\newblock Agbuilder: an ai tool for automated attack graph building, analysis,
  and refinement.
\newblock In {\em Data and Applications Security and Privacy XXXIII: 33rd
  Annual IFIP WG 11.3 Conference, DBSec 2019, Charleston, SC, USA, July 15--17,
  2019, Proceedings 33}, pages 23--42. Springer, 2019.

\bibitem{husari2017ttpdrill}
Ghaith Husari, Ehab Al-Shaer, Mohiuddin Ahmed, Bill Chu, and Xi~Niu.
\newblock Ttpdrill: Automatic and accurate extraction of threat actions from
  unstructured text of cti sources.
\newblock In {\em Proceedings of the 33rd annual computer security applications
  conference}, pages 103--115, 2017.

\bibitem{cve}
Cve.
\newblock \url{https://cve.mitre.org/}.

\bibitem{mitre}
The mitre corporation.
\newblock \url{https://www.mitre.org/}.

\bibitem{cve_repo}
Cve database.
\newblock \url{https://github.com/CVEProject/cvelistV5}.

\bibitem{huggingface}
Hugging face.
\newblock \url{https://huggingface.co/}.

\bibitem{jha2002two}
Somesh Jha, Oleg Sheyner, and Jeannette Wing.
\newblock Two formal analyses of attack graphs.
\newblock In {\em Proceedings 15th IEEE Computer Security Foundations Workshop.
  CSFW-15}, pages 49--63. IEEE, 2002.

\bibitem{noel2002combinatorial}
Steven~E Noel, Brian O'Berry, Charles Hutchinson, Sushil Jajodia, Lynn~M
  Keuthan, and Andy Nguyen.
\newblock Combinatorial analysis of network security.
\newblock In {\em Wavelet and Independent Component Analysis Applications IX},
  volume 4738, pages 140--149. SPIE, 2002.

\bibitem{ou2005mulval}
Xinming Ou, Sudhakar Govindavajhala, Andrew~W Appel, et~al.
\newblock Mulval: A logic-based network security analyzer.
\newblock In {\em USENIX security symposium}, volume~8, pages 113--128.
  Baltimore, MD, 2005.

\bibitem{hankin2022attack}
Chris Hankin, Pasquale Malacaria, et~al.
\newblock Attack dynamics: an automatic attack graph generation framework based
  on system topology, capec, cwe, and cve databases.
\newblock {\em Computers \& Security}, 123:102938, 2022.

\bibitem{jin2023prometheus}
Xin Jin, Charalampos Katsis, Fan Sang, Jiahao Sun, Elisa Bertino, Ramana~Rao
  Kompella, and Ashish Kundu.
\newblock Prometheus: Infrastructure security posture analysis with
  ai-generated attack graphs.
\newblock {\em arXiv preprint arXiv:2312.13119}, 2023.

\end{thebibliography}
